\documentclass[superscriptaddress, nofootinbib, twocolumn, amsmath,amssymb, aps, pra, notitlepage, longbibliography]{revtex4-1}

\usepackage{dcolumn}
\usepackage{bm}
\usepackage{epsfig}
\usepackage{graphicx}
\usepackage{latexsym}
\usepackage{amsfonts}
\usepackage{setspace}
\usepackage{graphicx}
\usepackage{amsmath}
\usepackage{verbatim}
\usepackage{color}
\usepackage{SIunits}
\usepackage{hyperref}
\usepackage{nccmath}
\usepackage{upgreek}

\begin{document}

\title{On-chip multi-stage optical delay based on cascaded Brillouin light storage}

\author{Birgit Stiller$^{1,\ast,\dagger}$, Moritz Merklein$^{1,\ast}$, Christian Wolff$^{2,3}$, Khu Vu$^{4}$,  Pan Ma$^{4}$, Christopher G. Poulton$^{2}$, Stephen J. Madden$^{4}$ and Benjamin J. Eggleton$^{1}$\\
\small{ \textcolor{white}{blanc\\}
$^{1}$Institute of Photonics and Optical Science (IPOS), The University of Sydney Nano Institute (Sydney Nano), School of Physics, The University of Sydney, NSW 2006, Australia.\\
$^{2}$School of Mathematical and Physical Sciences, University of Technology Sydney, NSW 2007, Australia.\\
$^{3}$Center for Nano Optics, University of Southern Denmark, Campusvej 55, DK-5230~Odense~M, Denmark.\\
$^{4}$Laser Physics Centre, RSPE, Australian National University, Canberra, ACT 2601, Australia.\\
$^{\ast}$These authors contributed equally to this work.\\
$^{\dagger}$birgit.stiller@sydney.edu.au}}

\begin{abstract} 

Storing and delaying optical signals plays a crucial role in data centers, phased array antennas, communication and future computing architectures. Here, we show a delay scheme based on cascaded Brillouin light storage, that achieves multi-stage delay at arbitrary positions within a photonic integrated circuit. Importantly these multiple resonant transfers between the optical and acoustic domain are controlled solely via external optical control pulses, allowing cascading of the delay without the need of aligning multiple structural resonances along the optical circuit.

\end{abstract}

\maketitle

The recent development of integrated photonic circuits benefits a range of applications, such as optical communication and signal processing \cite{Caucheteur2010,Lenz2001,Santagiustina2013}, phased array antennas \cite{Visser2005}, LIDAR \cite{Weitcamp2005}, optical and microwave delay lines \cite{Diehl2015,Liu2017,Chung2018,Wang2016} and future computing architectures that include photonic interconnects \cite{Alduino2007,Miller2009,Miller2010a}. In particular, with the rapid rise of integrated photonic circuits which allow signal processing in small-footprint devices, storing and delaying optical signals is crucial and on-chip delay schemes that can achieve large delays and bandwidth are indispensable. Long low loss waveguides can be fabricated to achieve a delay purely caused by the long physical path \cite{Lee2012a} but are not adjustable. Most on-chip approaches rely on resonant structures, such as rings \cite{Xie2017,Cardenas2010,Poon2004,Khurgin2009,Fiore2013} or photonic crystals \cite{Beggs2012,Nishikawa2002,Moreolo2008}, to achieve tunable delays in a small footprint, however the bandwidth is limited and the optical frequency is restricted to the resonances. Alternatively, stimulated Brillouin scattering (SBS) induced slow-light was implemented on chip \cite{Ravi2012}, however the fractional delay was limited to below unity. \newline
\begin{figure}[b!]
\centering
\includegraphics[width=0.93\linewidth]{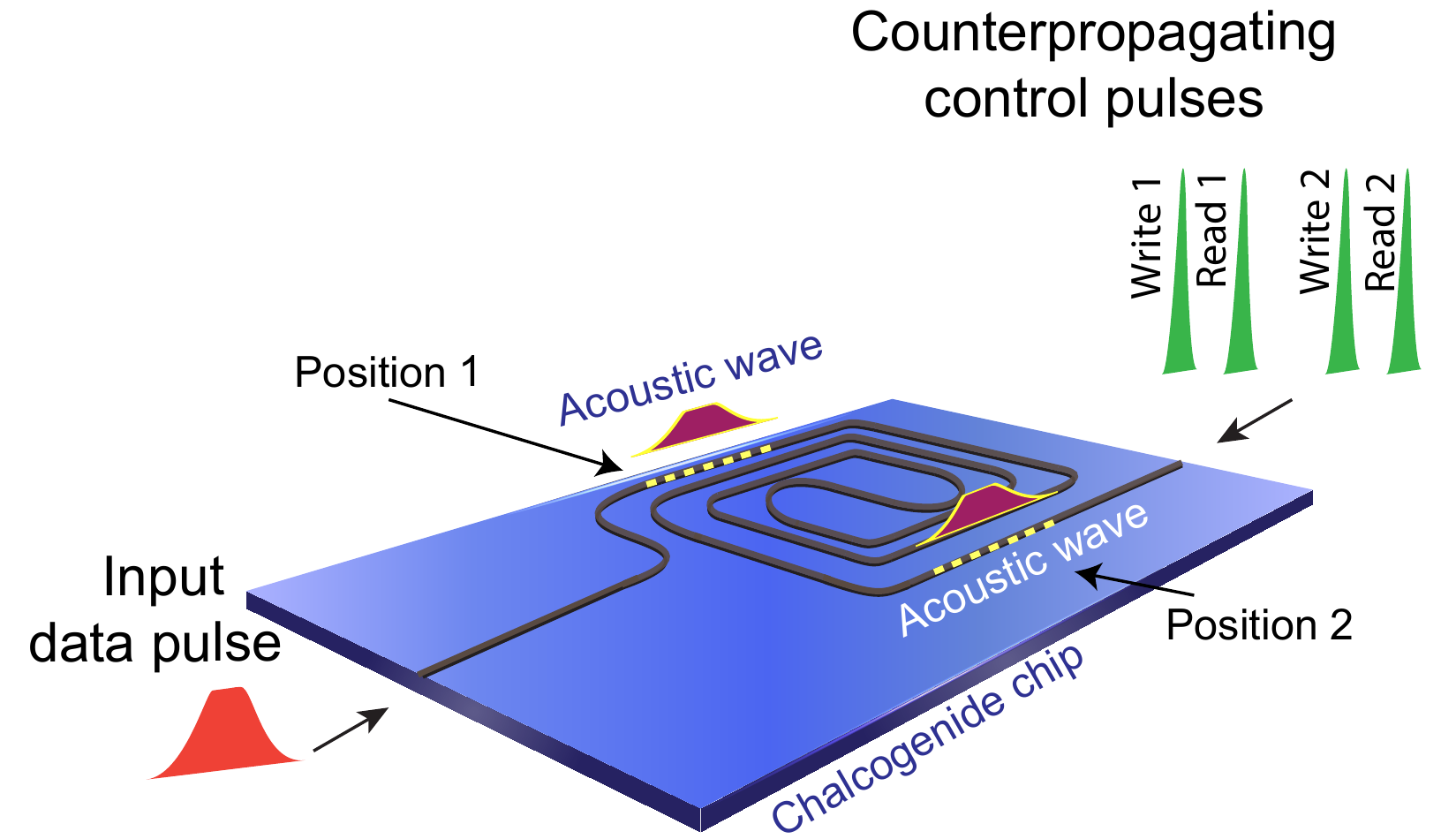}
\caption{Schematic of a planar spiral waveguide on an optical chip. The memory is operated at position 1 where the optical pulse is stopped for a first time and then, after it is transfered back to an optical signal is stopped a second time at position 2.}
\label{fig:1}
\end{figure}
\indent On the other hand it was recently shown that instead of reducing the group velocity of light a delay of an optical pulse can also be achieved by resonantly transferring it to an acoustic wave and retrieving it, within the constraints of the acoustic lifetime \cite{Zhu2007,Merklein2017}. This Brillouin based storage scheme allows for a large bandwidth, coherent storage and delay exceeding what is feasible using SBS slow-light schemes \cite{Merklein2017,Jaksch2017}. \newline
\indent Here, we report multi-stage operation of Brillouin-based light storage via cascading of the storage process. Within one photonic circuit light is transferred several times from the optical domain to the acoustic domain and vice versa. The whole process is governed solely by external optical control pulses and this allows for full spatial control over the storage process as we show light storage at multiple positions in the photonic circuit. With the multi-stage Brillouin-based light storage, we demonstrate the cascaded storage of an optical pulse two times, for 3\,ns each, leading to an overall delay of 6\,ns. Combining this multi-stage delay scheme with optical gain opens the possibility of reaching delay times beyond the acoustic lifetime. \newline
\indent The general idea of the multi-stage Brillouin memory is shown in an artist impression in Fig. \ref{fig:1}. A 1\,ns-data pulse interacts with a first pair of counter-propagating write pulses at position 1 in the photonic integrated circuit and is transferred to an acoustic wave. Afterwards the first optical read pulse is retrieving the optical pulse by transferring it back from the acoustic to the optical domain. The retrieved optical data pulse is traveling further in the spiral waveguide until it encounters, at interaction point 2, a second pair of write/read pulses and hence is transferred to an acoustic wave and back for a second time. The overall delay time is determined by the sum of the time difference \(t_1\) and \(t_2\) between the respective write and read pulses.

\begin{figure}[t!]
\centering
\includegraphics[width=\linewidth]{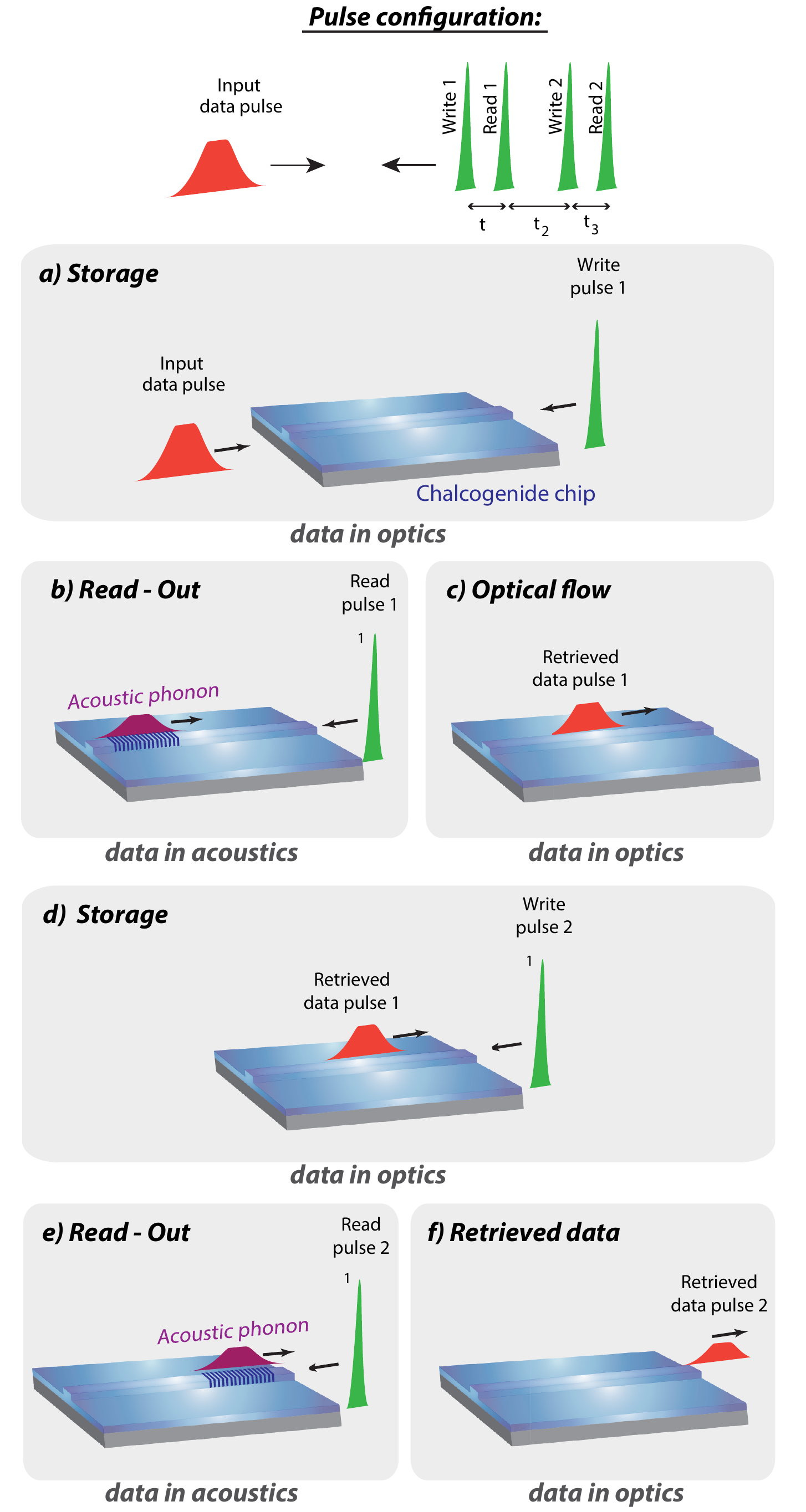}
\caption{Operation principle of the cascaded Brillouin memory. a) A data pulse and a first control pulse ``write 1'' are entering the waveguide. At the position the two counter-propagating pulses encounter each other, the data pulse is depleted and an acoustic wave is created. b) A second control pulse ``read 1'' enters the waveguide and is retrieves the optical data pulse from the acoustic wave while depleting the later. c) A first read-out data pulse is traveling further in the optical domain. d) A third pulse, ``write 2'', stores the optical pulse for a second time as an acoustic wave at position 2. e) A fourth pulse, ``read 2'', reads out the acoustic wave. f) A data pulse, delayed by sum of the individual storage times as acoustic phonons, exits the waveguide.}
\label{fig:2}
\end{figure}

\begin{figure}[t!]
\centering
\includegraphics[width=\linewidth]{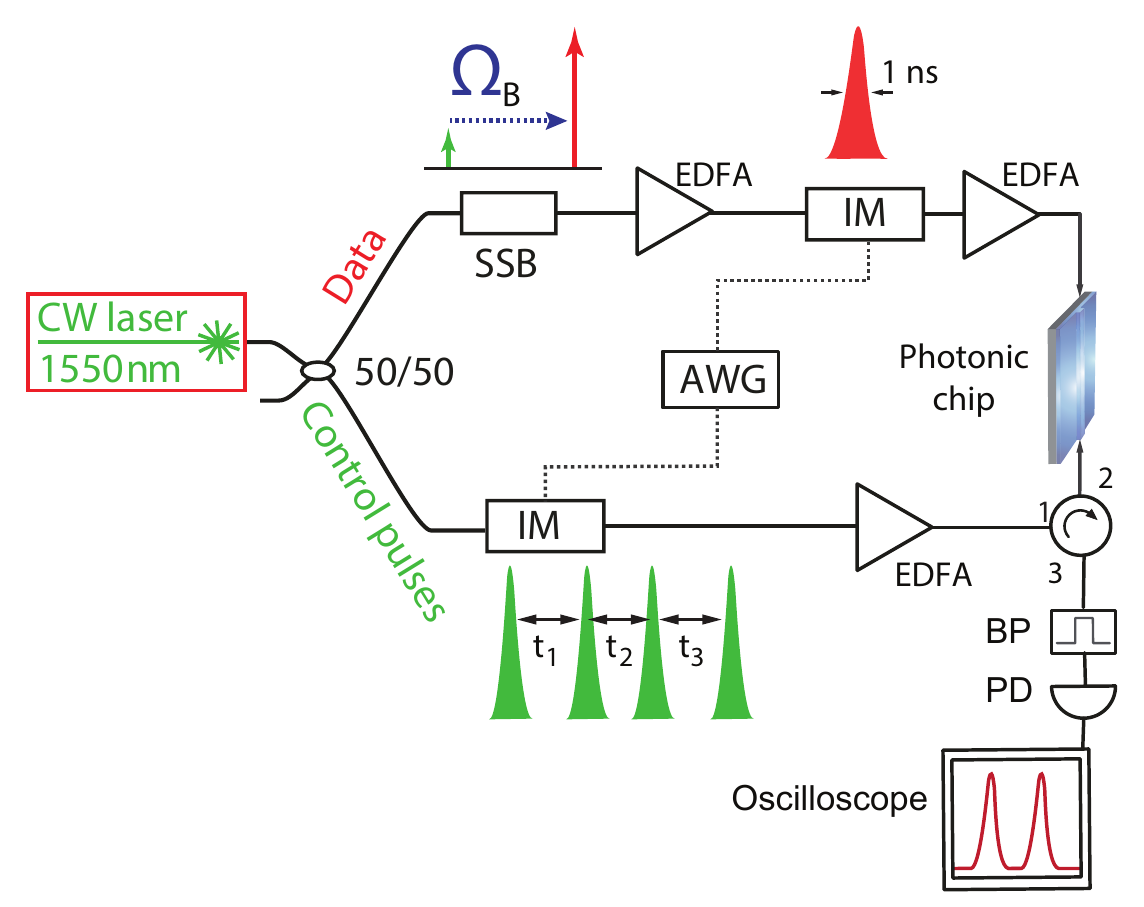}
\caption{Experimental setup. SSB single sideband modulator, EDFA erbium doped fiber amplifier, IM intensity modulator, AWG arbitrary waveform generator, BP bandpass filter, PD photo diode} 
\label{fig:setup}
\end{figure}

\begin{figure}[t!]
\centering
\includegraphics[width=\linewidth]{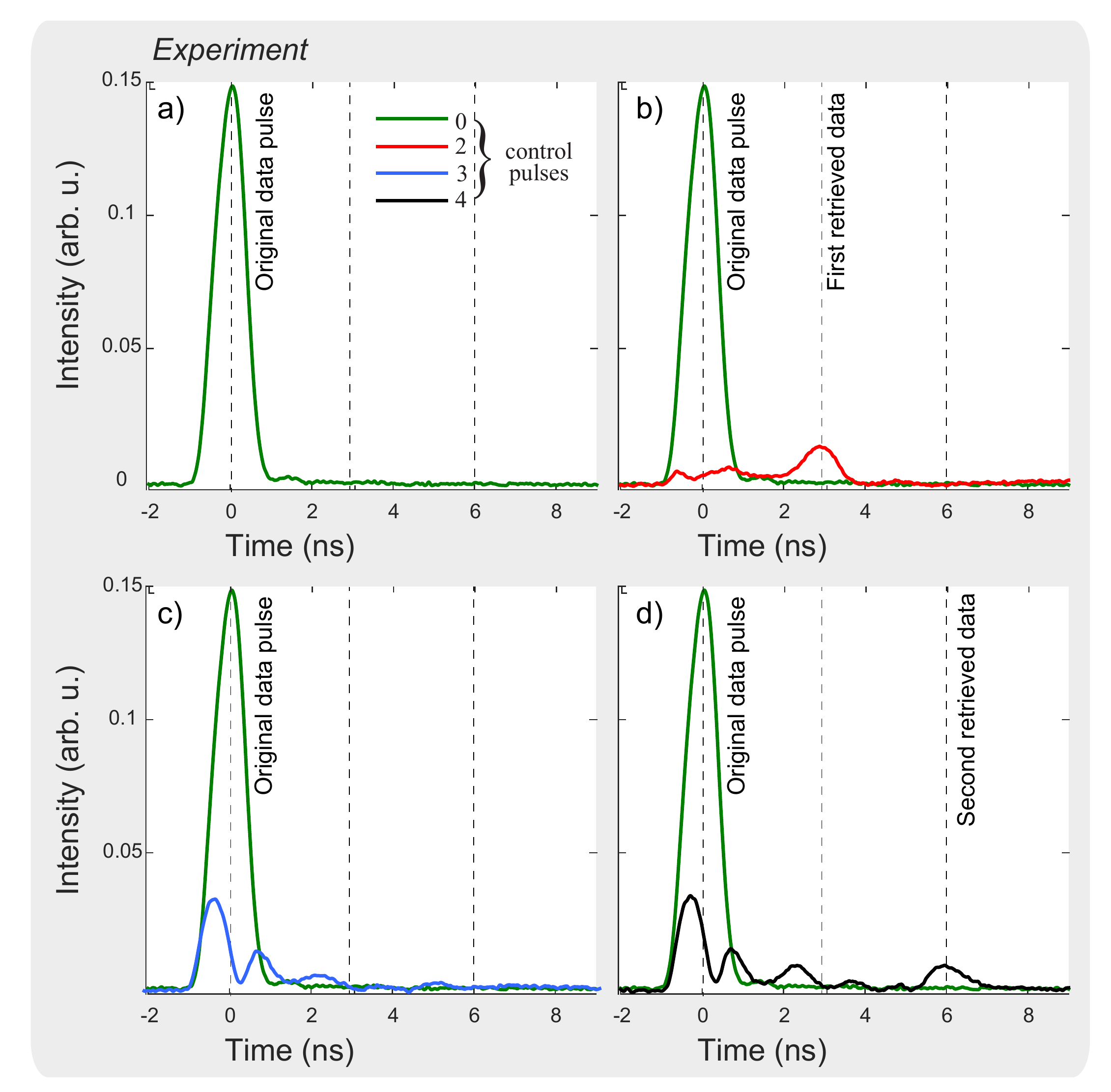}
\caption{Experimental results of the cascaded memory process. a) Original data pulse. b) Read-out after 3\'ns using two control pulses ``write 1'' and ``read 1''. c) Depletion of the retrieved optical data pulse by operating the memory with three control pulses: ``write 1'', ``read 1'' and ``write 2''. d) Second read-out after 6\,ns after a double storage process with pulses ``write 1'', ``read 1'', ``write 2'' and ``read 2''.}
\label{fig:4}
\end{figure}

\begin{figure}[t!]
\centering
\includegraphics[width=\linewidth]{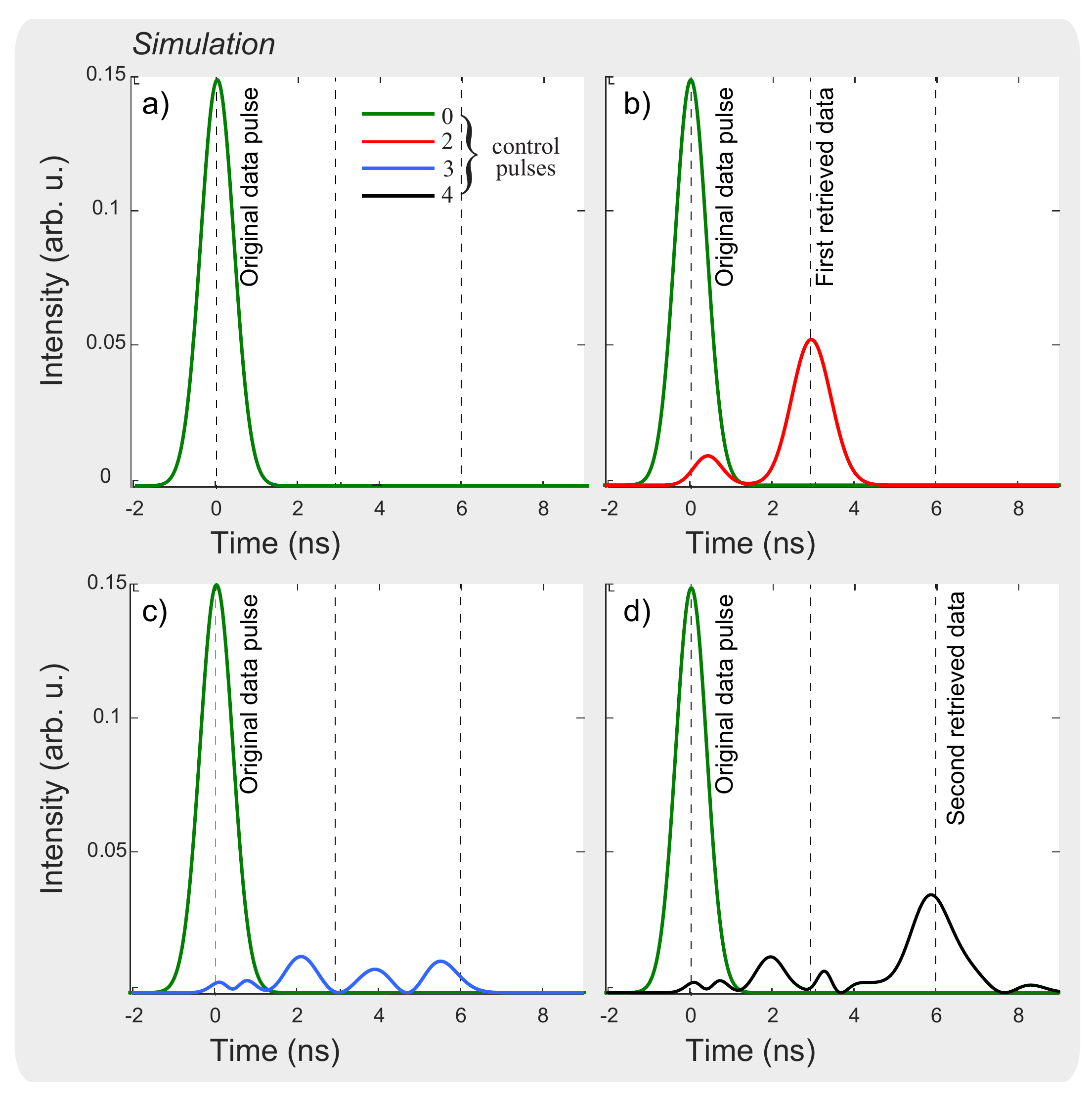}
\caption{Simulation of the cascaded memory process. a) Original data pulse. b) Read-out after 3\'ns using two control pulses ``write 1'' and ``read 1''. c) Depletion of the read-out by operating the memory with three control pulses: ``write 1'', ``read 1'' and ``write 2''. d) Second read-out after 6\,ns after a double storage process with pulses ``write 1'', ``read 1'', ``write 2'' and ``read 2''.}
\label{fig:3}
\end{figure}

\indent A more detailed view on the different stages in terms of pulse propagation and the transfer between the optical and acoustic domain is given in Fig.~\ref{fig:2}. The optical data pulse with carrier frequency $\omega_{\text{data}}$ is coupled into the waveguide from the left. This data pulse encounters a train of four counter-propagating control pulses separated by time intervals $t_1=3\,\text{ns}$, $t_2=2\,\text{ns}$ and $t_3=3\,\text{ns}$. The control pulses' carrier frequency $\omega_{\text{control}}$ is offset from the data carrier by the Stokes shift $\Omega_{\text{B}} = \omega_{\text{data}} - \omega_{\text{control}}$.  Within the control pulse train, pulses 1\,\&\,3 serve as ``write'', and pulses 2\,\&\,4 as ``read'' pulses. The first ``write'' pulse converts the data to a coherent acoustic wave via the Stokes-process of SBS (see Fig.~\ref{fig:2}a,b). The data is then back-converted to the optical domain by the first ``read'' pulse via the anti-Stokes process (see Fig.~\ref{fig:2}b). The information travels some distance as an optical pulse (Fig.~\ref{fig:2}c) before the ``write''--``read'' cycle is repeated (see Figs.~\ref{fig:2}d,e\,\&\,f). This double process delays the data pulse by the time $t_1 + t_3$ with respect to free propagation. Spurious interactions between data and control pulses and reflections thereof can create signals at other combinations of $t_1$, $t_2$ and $t_3$, but the double-encoding process is the only way to obtain a signal at $t_1 + t_3$. Our values for $t_1$, $t_2$ and $t_3$ allow us to unambiguously separate this from any spurious response and thus demonstrate the repeated encoding and decoding of optical information as an acoustic signal. \newline
\indent The storage medium used for the on-chip demonstration of the cascaded opto-acoustic light storage is a chalcogenide glass $As_{2}S_{3}$ waveguide. The planar rib waveguide structure has a cross-section of 2.2 $\mathrm{\upmu}$m by 800\,nm and a total length of 46\,cm and is arranged as a spiral on the photonic chip to ensure a small footprint. Details on the fabrication methods of the chip can be found in Ref.~\cite{Madden2007}. Record SBS gain could be observed in this type of waveguides with a specific design that embeds the chalcogenide glass between a silica substrate and a silica over cladding which provides a large opto-acoustic overlap \cite{Choudhary2017}. \newline
\indent The experimental setup is depicted in Fig. \ref{fig:setup}. A narrow linewidth laser at 1550\,nm is used as a source for both data and the control pulses. In the data arm, the laser is first up-shifted by the Brillouin frequency shift $\Omega_{B}$, in our case 7.65\,GHz for the chalcogenide waveguide, and then sent through an intensity modulator that creates a 1\,ns long pulses using an arbitrary waveform generator (AWG). Two low-noise erbium doped fiber amplifiers (EDFA) compensate for induced losses. A polarization controller before the chip adjusts the polarization such that data and control pulses interact efficiently as SBS is polarization dependent. In the control arm, four control pulses are created by the second channel of the AWG and are pre-amplified by an EDFA. The pulses then pass through a nonlinear loop (not depicted in the setup) which reduces the noise floor and gives us additional flexibility in terms of pulse shaping. Before entering the photonic waveguide from the opposite side, the control pulses are amplified by an EDFA and the polarization is adjusted. After passing through the chip and encountering the four control pulses, the data pulse is filtered by a narrow linewidth tunable filter, in order to prevent a spurious signal from back-reflection of the control pulses which are at a different frequency than the data pulses. The pulses are detected by an electrically amplified photo detector and observed at a 12\,GHz oscilloscope. \newline
\indent Figure \ref{fig:4} shows the operation of the photonic memory for a different number of control pulses. In the absence of any control pulses, the output only consists of the original data pulse (green in Fig. \ref{fig:4}a). For comparison, the original data is superimposed in Fig. \ref{fig:4}b-d. When coupling two control pulses into the waveguide, the original data pulse is depleted and a read-out can be observed that is delayed by 3\,ns relative to the free propagation through the waveguide (Fig. \ref{fig:4}b). Note that the power of the control pulses have to be precisely adjusted such that the ``write 1'' pulse completely depletes the original data pulse and the ``read 1'' pulse reads out the acoustic wave efficiently such that we do not observe any spurious reflections or unwanted storage operations. For example, in case of the original data pulse not being completely depleted, the second control pulse can act as another write pulse. \newline
\indent Switching the memory operation from two to three pulses (Fig. \ref{fig:4}c), allows for a storage-retrieval-storage process. Therefore the original data pulse is encoded and retrieved at position 1 (compare Fig. \ref{fig:1}) once, travels further in the optical domain until it encounters the third control pulse ``write 2''. As a result, the data pulse is depleted for a second time and the information is transferred to a second acoustic wave at position 2 in the photonic circuit. Hence, in Fig. \ref{fig:4}c, there is no read-out observed. When coupling a fourth pulse (3\,ns delayed) in the waveguide, it reads out the acoustic wave created at position 2 and we obtain a read-out 6\,ns after the original data pulse. As mentioned above, the optical power levels have to be adjusted carefully, such that the pulses act as alternating write and read pulses. \newline
\indent We simulated the cascaded memory using standard coupled mode equations \cite{Wolff2015,Winful2016} and an implicit fourth order Runge-Kutta approach as a solver \cite{Winful2016}. The results of the numerical study are in qualitative agreement with our experimental results and are presented in Fig. \ref{fig:3}a-d. We see a first readout after \(t_1\) for two control pulses, a second depletion of the retrieved data in the three control pulse configuration and a readout after \(t_1 + t_2\) for four counter-propagating control pulse. The coupled mode equations for the optical waves (data and write/read) and the acoustic wave under the assumption of a slowly-varying envelope read as:
\begin{equation}
\frac{\partial A_{\mathrm{D}}}{\partial z} + \frac{n}{c} \frac{\partial A_{\mathrm{D}}}{\partial t}  = -\frac{g_{\mathrm{0}}}{2 A_{\mathrm{eff}}} Q A_{\mathrm{C}} - \frac{1}{2} \alpha A_{\mathrm{D}}
\end{equation}
\begin{equation}
- \frac{\partial A_{\mathrm{C}}}{\partial z} + \frac{n}{c} \frac{\partial A_{\mathrm{C}}}{\partial t}  = \frac{g_{\mathrm{0}}}{2 A_{\mathrm{eff}}} Q^{*} A_{\mathrm{D}} - \frac{1}{2} \alpha A_{\mathrm{C}}
\end{equation}
\begin{equation}
2 \tau_{\mathrm{B}} \frac{\partial Q}{\partial t} + Q = A_{\mathrm{D}} A_{\mathrm{C}}^{*}
\end{equation}


where $A_{\mathrm{D}}$ is the data pulse, $A_{\mathrm{C}}$ the counter propagating control pulses (write and read) and $Q$ the traveling acoustic wave. The other parameters are the Brillouin gain coefficient $g_{\mathrm{0}}$, the acoustic lifetime $\tau_{\mathrm{B}}$, the linear loss of the waveguide $\alpha$, the speed of light $c$ and the effective refractive index $n$. The amplitudes of the optical and acoustic waves have been normalized such that $|A_{\mathrm{D/C}}|^{2}$ is the power in Watts and the data and control pulses have a Gaussian shape \cite{Winful2016}:
\begin{equation}
A_{\mathrm{D/C}} = A_{0} \mathrm{exp} \left(- \frac{1+i C}{2} \frac{t^{2}}{\tau^{2}}\right)
\end{equation}
with the parameter C giving the chirp rate in GHz\,/\,ns following the definition of \cite{Winful2016} and $\tau$ being the FWHM. \newline
\indent In conclusion we have demonstrated a multi-stage delay scheme implementation based on Brillouin light storage, that relies on the storage of optical information as acoustic waves. Cascaded operation is generally hard to achieve in other delay schemes that are based on structural resonances, as one has to make sure that the multiple resonances are spectrally aligned. We demonstrated the flexibility of Brillouin light storage that does not rely on a structural resonance but can be implemented in a planar waveguide. Multiple transfers between an acoustic and optical wave within a single waveguide was shown by sending in a cascade of write and read pulses. These multiple transfers are completely determined by the shape and temporal separation of the external control pulses, which does not require any intervention between the transfers and hence show the great control of the coherent transfer between acoustic and optical waves that can be achieved. Besides the full spatial control of the storage/delay operation within a photonic circuit, the cascaded storage scheme opens the path for extended temporal control in Brillouin based, or in a broader context, phonon based light storage or delay. Combining the cascaded operation with optical gain will allow an extension of the storage time beyond the acoustic lifetime, that is so far a limiting factor in phonon-based light storage.

\section*{Funding Information}
This work was sponsored by the Australian Research Council (ARC) Laureate Fellowship (FL120100029) and the Centre of Excellence program (CUDOS CE110001018). We acknowledge the support of the ANFF ACT.








\begin{thebibliography}{10}




 \bibitem{Caucheteur2010}
C. Caucheteur, A. Mussot, S. Bette, A. Kudlinski, M. Douay, E. Louvergneaux, P. M\'{e}gret, M. Taki, and Miguel Gonz\'{a}lez-Herr\'{a}ez, Opt. Express \textbf{18}, 3093-3100 (2010).

 \bibitem{Lenz2001}
G. Lenz, B. Eggleton, C. K. Madsen, and R. Slusher, IEEE Journal of Quantum Electronics \textbf{37}, 525--532 (2001).

 \bibitem{Santagiustina2013}
M. Santagiustina, S. Chin, N. Primerov, L. Ursini, and L. Th{\'{e}}venaz, Sci. Rep. \textbf{3}, doi:10.1038/srep01594 (2013).

 \bibitem{Visser2005}
H. J. Visser, ``Array and Phased Array Antenna Basics,'' John Wiley \& Sons, DOI 10.1002/0470871199 (2005).

 \bibitem{Weitcamp2005}
C. Weitkamp, ``Lidar - Range-Resolved Optical Remote Sensing of the Atmosphere,'' Springer Series in Optical Sciences, Springer-Verlag New York, DOI 10.1007/b106786 (2005).

  \bibitem{Diehl2015}
J. F. Diehl, J. M. Singley, C. E. Sunderman, and V. J. Urick, Appl. Opt.  \textbf{54}, F35--F41 (2015).

\bibitem{Liu2017}
Y. Liu, A. Choudhary, D. Marpaung, and B. J. Eggleton, Optica \textbf{4}, 418-423 (2017).

\bibitem{Chung2018}
C.-J. Chung, X. Xu, G. Wang, Z. Pan, and R. T. Chen, Appl. Phys. Lett. \textbf{112}, 071104 (2018).

\bibitem{Wang2016}
J. Wang, R. Ashrafi, R. Adams, I. Glesk, I. Gasulla, J. Capmany, and Lawrence R. Chen, Scientific Reports \textbf{6}, 30235, doi:10.1038/srep30235 (2016).

  \bibitem{Alduino2007}
A.~Alduino and M.~Paniccia, Nature Photonics \textbf{1}, 153--155 (2007).

\bibitem{Miller2009}
D.~Miller, Proceedings of the IEEE \textbf{97}, 1166--1185 (2009).

  \bibitem{Miller2010a}
D.~Miller, Applied Optics \textbf{49}, 70 (2010).

  \bibitem{Lee2012a}
H.~Lee, T.~Chen, J.~Li, O.~Painter, and K.~J. Vahala, Nature Communications \textbf{3}, 867 (2012).
  
\bibitem{Xie2017}
Y. Xie, L. Zhuang, K.-J. Boller, and A. J. Lowery, Journal of Optics \textbf{19}, 6 (2017).

\bibitem{Cardenas2010}
J. Cardenas, M. A. Foster, N. Sherwood-Droz, C. B. Poitras, H. L. R. Lira, B. Zhang, A. L. Gaeta, J. B. Khurgin, P. Morton, and M. Lipson, Opt. Express \textbf{18}, 26525-26534 (2010).
   
\bibitem{Poon2004}
J. K. S. Poon, J. Scheuer, Y. Xu, and A. Yariv, J. Opt. Soc. Am. B \textbf{21}, 1665-1673 (2004). 

\bibitem{Khurgin2009}
J. B. Khurgin and P. A. Morton, Opt. Lett. \textbf{34}, 2655-2657 (2009).

\bibitem{Fiore2013}
V. Fiore, C. Dong, M. C. Kuzyk, and H. Wang, Phys. Rev. A \textbf{87}, 1--6 (2013).

\bibitem{Beggs2012}
D. M. Beggs, I. H. Rey, T. Kampfrath, N. Rotenberg, L. Kuipers, and T. F. Krauss, Phys. Rev. Lett. \textbf{108}, 213901 (2012).

\bibitem{Nishikawa2002}
S. Nishikawa, S. Lan, N. Ikeda, Y. Sugimoto, H. Ishikawa, and K. Asakawa, Opt. Lett. \textbf{27}, 2079-2081 (2002).

\bibitem{Moreolo2008}
M. Svaluto Moreolo, V. Morra and G. Cincotti, Journal of Optics A: Pure and Applied Optics \textbf{10}, 6 (2008).

\bibitem{Ravi2012}
R. Pant, A. Byrnes, C. G. Poulton, E. Li, D.-Y. Choi, S. Madden, B. Luther-Davies, and B. J. Eggleton, Opt. Lett. \textbf{37}, 969-971, (2012).

\bibitem{Zhu2007}
Z. Zhu, D. J. Gauthier, and R. W. Boyd, Science \textbf{318}, 1748–50 (2007).

\bibitem{Merklein2017}
M. Merklein, B. Stiller, K. Vu, S. J. Madden, and B. J. Eggleton, Nature Comm. \textbf{8}, 574, doi:10.1038/s41467-017-00717-y (2017), arXiv ID: 1608.08767.

\bibitem{Jaksch2017}
K. Jaksch, M. Merklein, K. Vu, P. Ma, S. J. Madden, B. J. Eggleton, and B. Stiller, Frontiers in Optics/Laser Science Conference (FiO/LS) 2017, Washingtion D.C., USA, postdeadline FTh4A.5
  
\bibitem{Madden2007}
S. J. Madden, D-Y. Choi, D. A. Bulla, A. V. Rode, B. Luther-Davies, V.G. Ta’eed, M.D. Pelusi, and B.J. Eggleton,  Opt. Express \textbf{15}, 14414-14421 (2007).

\bibitem{Choudhary2017}
 A. Choudhary, B. Morrison, I. Aryanfar, S. Shahnia, M. Pagani, Y. Liu, K. Vu, S. Madden D. Marpaung, B. J. Eggleton, IEEE J. Light. Technol. \textbf{35}, 4 (2017).
  
\bibitem{Wolff2015}
C.~Wolff, M.~J. Steel, B.~J. Eggleton, and C.~G. Poulton, Phys. Rev. A \textbf{92}, 013836 (2015).

\bibitem{Winful2016}
  M.~Dong, and H.~G. Winful, Phys. Rev. A \textbf{93}, 043851 (2016).

\end{thebibliography}


\end{document}